\documentclass[pdflatex,runningheads]{llncs}
\usepackage[T1]{fontenc}
\usepackage{graphicx}

\usepackage{graphicx}
\usepackage{algpseudocode}
\usepackage{amsmath,amssymb,amsfonts}
\usepackage{subfigure}
\usepackage{subcaption}
\usepackage{caption}
\usepackage{ulem} 
\usepackage{booktabs}
\usepackage{tabularx}
\usepackage{xcolor} 
\usepackage{enumitem}
\usepackage{multirow} 
\usepackage{calc}
\usepackage{pifont}
\usepackage{wrapfig}

\usepackage{microtype}
\usepackage{float}
\usepackage{algorithm}
\usepackage{hyperref}
\usepackage[]{hyperref}
\hypersetup{
        colorlinks=true,
        citecolor=blue,
	linkcolor=blue,
	urlcolor=blue,
	citecolor=blue,
}

\usepackage{color}

\begin{document}
%

\title{SpecInF: Exploiting Idle GPU Resources in Distributed DL Training via Speculative Inference Filling}

%
%
%
\author{Cunchi Lv\inst{1,2,4} \and
Xiao Shi\inst{1,5} \and
Dong Liang\inst{1} \and
Wenting Tan\inst{1} \and
Xiaofang Zhao \inst{1,3,4}
}

\authorrunning{Cunchi Lv, Xiao Shi, Dong Liang, Wenting Tan, Xiaofang Zhao}
\titlerunning{SpecInF: Exploiting Idle GPU Resources in Distributed DL Training }

%
%
\institute{Institute of Computing Technology, Chinese Academy of Sciences \and
University of Chinese Academy of Sciences \and
University of Chinese Academy of Sciences, Nanjing \and
Zhongguancun Laboratory \and
Nanjing Institute of InforSuperbahn  \\
\email{\{lvcunchi21s,shixiao,liangdong,tanwenting,zhaoxf\}@ict.ac.cn}}
\maketitle              

\begin{abstract}

Deep Learning (DL), especially with Large Language Models (LLMs), brings benefits to various areas.
However, DL training systems usually yield prominent idling GPU resources due to many factors, such as resource allocation and collective communication.
To improve GPU utilization, we present SpecInF, which adopts a \textbf{Spec}ulative \textbf{In}ference \textbf{F}illing method to exploit idle GPU resources. It collocates each primary training instance with additional inference instances on the same GPU, detects the training bubbles and adaptively fills with online or offline inference workloads.
Our results show that SpecInF can effectively enhance GPU utilization under mainstream parallel training modes, delivering additional up to 14$\times$ offline inference throughputs than TGS and 67\% reduction in online inference p95 latency than MPS, while guaranteeing collocated training throughput.



\end{abstract}
\section{Introduction}



The rapid progress in deep learning (DL) has significantly benefitted various areas, like manufacturing \cite{manufacture}, artistic creation \cite{art}, and online services \cite{ChatGPT}, especially with the emergence of the Large Language Models (LLMs). For example, ChatGPT \cite{ChatGPT} facilitates a remarkable breakthrough in this evolution. Alongside this, the growth in LLM training has led to a surge in demand for GPUs, with substantially considerable costs. For example, OpenAI used approximately 25,000 Nvidia A100 GPUs for about 90 to 100 days to train GPT-4 \cite{training-GPT}, costing around 63 million dollars. The cost efficiency is more non-negligible than ever.

However, GPU utilization in DL training is considerably low due to many factors in distributed patterns, such as well-known communication overheads \cite{xiao2020antman}, which largely pulls down the cost efficiency.
Although various distributed training patterns greatly shorten the training time, they can also cause GPU resource wastage. For instance, when training GPT-4, the average GPU utilization only ranged between 32\% and 36\% \cite{training-GPT}.
This low usage is primarily due to communication overhead, despite significant optimizations in frameworks(e.g., DeepSpeed \cite{DeepSpeed}, Colossal \cite{ColossalAI}, NVIDIA Megatron \cite{shoeybi2019megatron}, and PyTorch.DDP \cite{DDP}).
Thus, instead of directly optimizing the training workflow, we argue that the idling GPU resources can be reassigned to serve DL inferences.
We observe that there exists a complementarity in both memory and compute resource consumption between small- or medium-sized inference and distributed training. Therefore, it can significantly improve GPU utilization by filling DL training idling phases with inference workloads. 

In this paper, we present \textbf{SpecInF}, a system that leverages a \textbf{Spec}ulative \textbf{In}ference \textbf{F}illing mechanism, to exploit idling GPU resources of distributed training and increase aggregated throughputs of GPUs.
First, it allows inference instances to collocate with training instances according to their memory demands and GPU idling characteristics. Second, it adopts a Bubble Monitor to detect GPU idling timing in real time. Third, it builds a CUDA Kernel Scheduler to issue tokens to collocated inference instances, in which the Kernel Barrier decides to release inference CUDA kernels to fill training bubbles.
In summary, our contributions are as follows:
\begin{itemize}
    \item We analyze the GPU fragmentation in distributed training, especially for LLMs, and propose to collocate it with inference instances to improve the utilization.
    \item We design the speculative inference filling mechanism, allowing adaptive kernel scheduling to efficiently serve both online and offline inferences.
    \item We build and evaluate the SpecInF system. The experiments show that SpecInF significantly improves GPU utilization of various distributed training modes, delivering additional up to 14$\times$ offline inference throughputs than TGS and 67\% reduction in online inference p95 latency than MPS, while guaranteeing training throughput.

\end{itemize}

\section{Background and Motivation}
\setcounter{footnote}{0}

\subsection{Distirubted DL Training and Idle GPU Resources}
\label{sec:opportunity}

Distributed DL training has been widely used to accelerate and improve throughput by utilizing multiple GPUs in parallel, mainly including Data Parallelism (DP) \cite{li13pytorch}, Model Parallelism (MP) \cite{shoeybi2019megatron}, Pipeline Parallelism (PP) \cite{narayanan2019pipedream}, and Hybrid Parallelism (HP) \cite{hp}. 
The parallel strategies divide datasets (e.g., DP) or models (e.g., MP, PP) into multiple GPUs, and complete the forward and backward propagation with explicit communication among GPUs. 
However, the GPUs of training clusters are usually underutilized due to many factors, including unreasonable resource allocation, communication overhead and failure recovery. Consequently, idle GPU resources may exist on both compute and memory aspects.

\begin{figure*}[h] 
    \centering
    \setlength{\abovecaptionskip}{-0.cm}
    \subfigure[DP mode]{
            \includegraphics[scale=0.76]{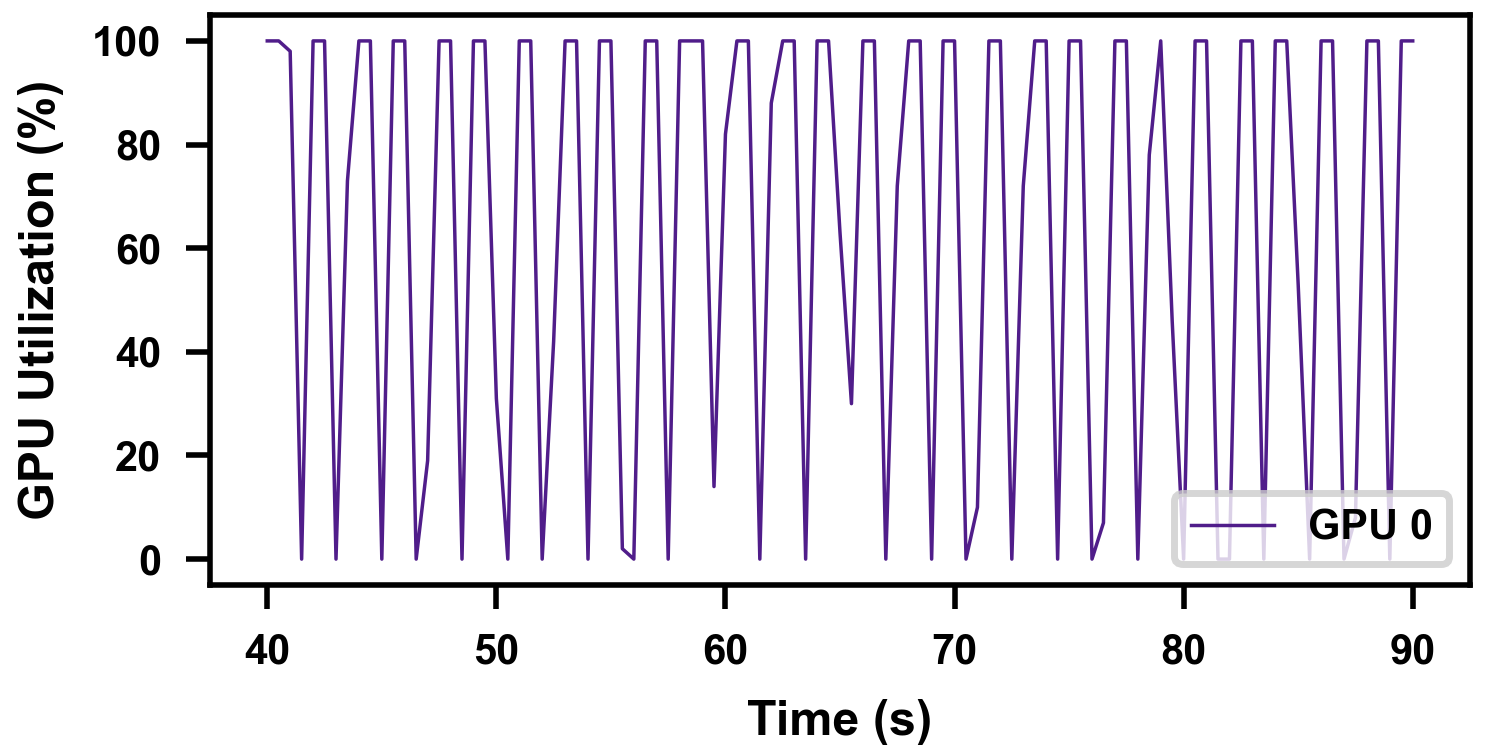}
		\label{fig:ddp-profiling}
    }
    \hspace{-0.1 in}
    \subfigure[MP mode]{
            \includegraphics[scale=0.76]{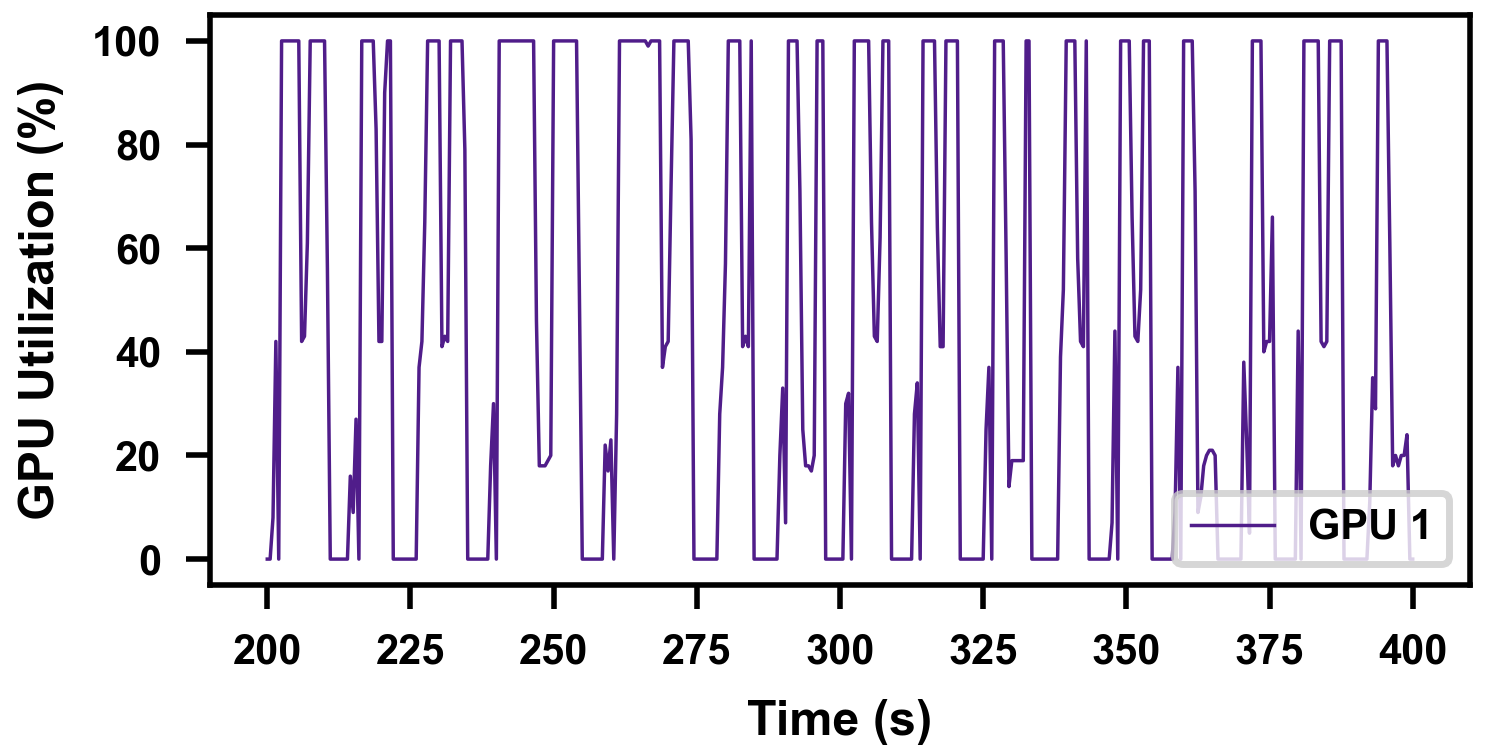}
            \label{fig:mp-profiling}
    }
	
\caption{The GPU compute utilization timeline of two modes, as monitored by the nvml APIs. (a) training RoBERTa-large model in DP mode via PyTorch.DDP; (b) fine-tuning LLaMA2-7B in MP mode via DeepSpeed. Both two cases involve 4 GPU workers. }
	\label{fig:profiling-performance}
\vspace{-0.1in}
\end{figure*}

\begin{figure*}[h] 
    \centering
    \setlength{\abovecaptionskip}{-0.cm}
    \subfigure[GPU consumption on training]{
            \includegraphics[scale=0.77]{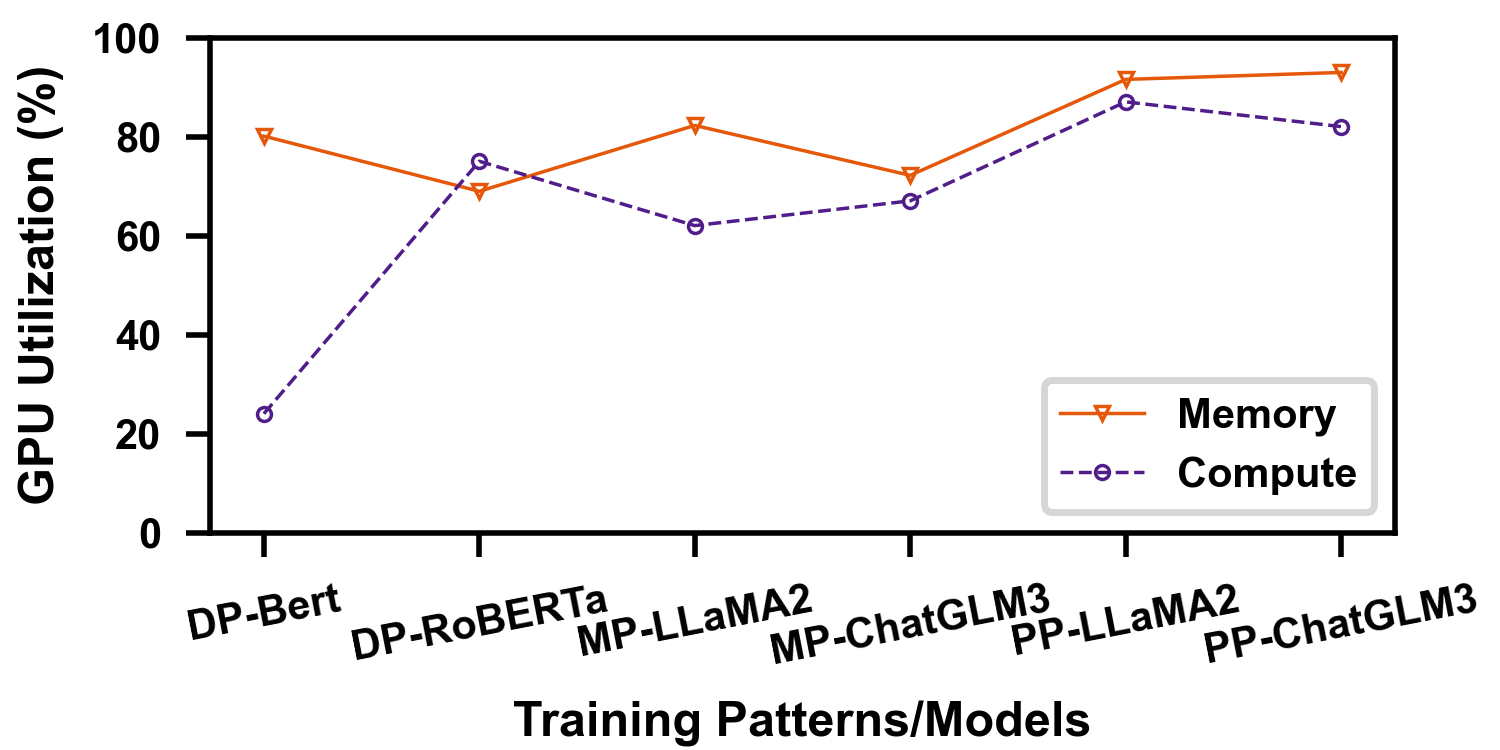}
		\label{fig:training-profiling}
    }
    \hspace{-0.1 in}
    \subfigure[GPU consumption on inference]{
     	\includegraphics[scale=0.77]{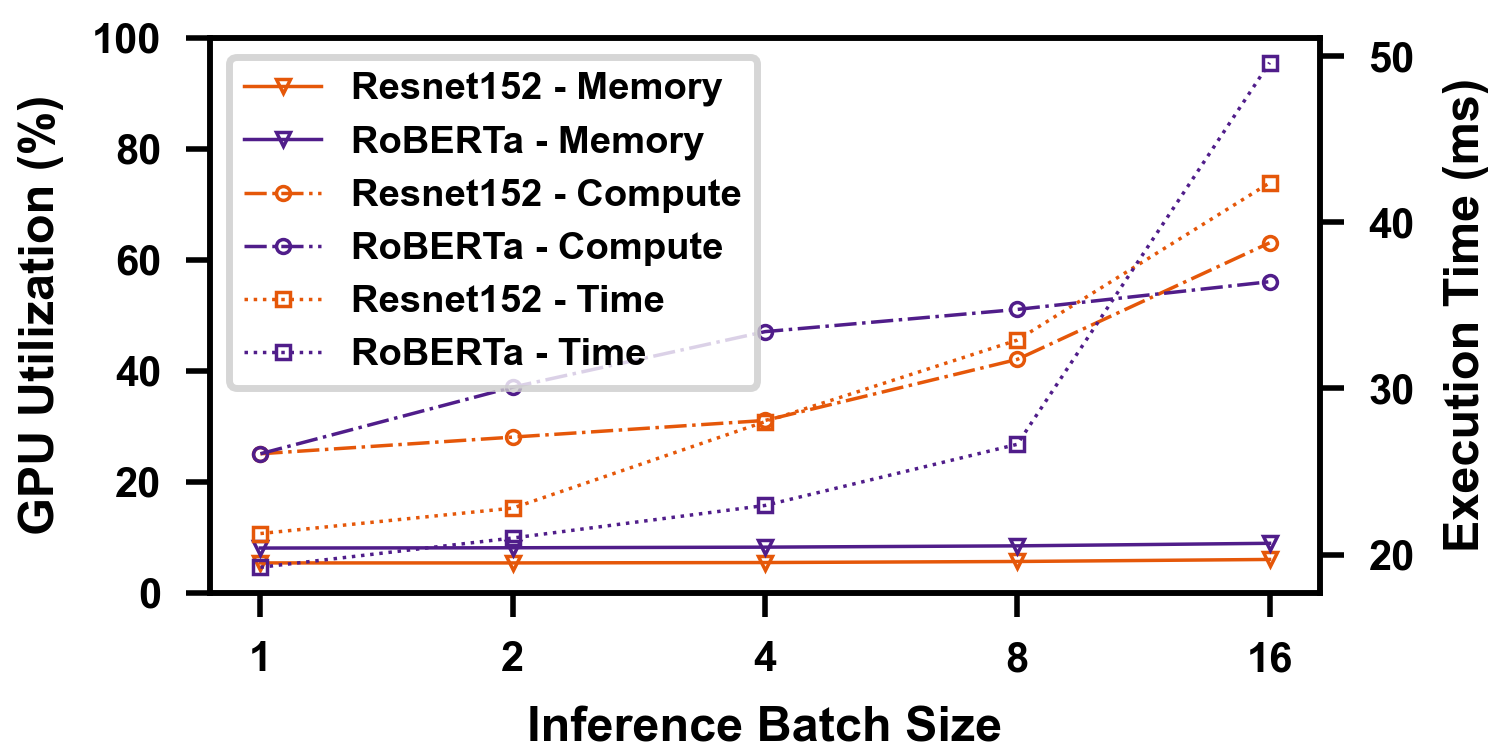}
            \label{fig:inference-profiling}
    }
	
\caption{GPU occupying characteristics of distributed training and inference.}
	\label{fig:training-inference-profiling}
\vspace{-0.2in}
\end{figure*}


\textbf{Idle GPU Computing Resources}. The communication among GPUs causes explicit temporal idling in GPU computing resources. Despite the optimizations (e.g., torch.DDP \cite{DDP}, DeepSpeed \cite{DeepSpeed}, and ColossalAI \cite{ColossalAI}), it is still hard to fully overlap computation with communication. Figure~\ref{fig:ddp-profiling} illustrates that there is nearly 30\% GPU-time is waiting for communication in the DP mode. Figure~\ref{fig:mp-profiling} depicts that GPU utilization of MP shows similar situations. Furthermore, we observe that the average compute utilization across various training tasks ranges from 20\% to 80\%, as shown in the purple line of Figure~\ref{fig:training-profiling}.

\textbf{GPU Memory Fragments}. Although training tasks are typically memory-bound \cite{xiao2020antman,wu2023transparent,chatglm3,llama2}, there exist various GPU memory fragments in practice, since the local batch size is often set with powers of 2, aiming to ensure training convergence and maximize GPU performance \cite{masters2018revisiting,park2022efficient}. As illustrated by the orange line in Figure~\ref{fig:training-profiling}, an average memory fragmentation of 10-20\% across three training modes is observed. For the GPU with 40 GB memory, the idle memory resource can be 4 to 8 GB.

\subsection{Insight and Challenges}


\textbf{Insight}. Considering the temporal (compute resource) and spatial (memory resource) fragments of GPUs, we argue that these unused resources can be adequate to serve moderate DL inference workloads.
As shown in Figure~\ref{fig:inference-profiling}, with the inference batch size increasing, the memory resource consumption stays stable due to the unnecessity of storing gradients and activations, and built-in memory pooling mechanism \cite{li13pytorch}, while the utilization of GPU compute resources notably increases, by 25\% to 56\% of RoBERTa. The corresponding inference latencies (at the 50ms level) are much shorter than the training iteration (as shown in Figure~\ref{fig:ddp-profiling}).
Thus, we propose a speculative inference filling method to improve the GPU utilization of specific training clusters, allowing to reassign idling GPU resources to handle moderate inference workloads.


\textbf{Goals and Challenges}. To improve the GPU utilization of training clusters, we attempt to build a speculative inference filling mechanism integrated into the training system. This mechanism aims to \emph{maintain the training throughput while concurrently providing additional inference services}.  The challenges include two aspects. First, it is difficult to decide the speculative filling timing. Second, it is necessary to avoid interference between the training and inference workloads. The timing accuracy and interference may affect the performance of both training and inference workloads.
\label{Sec:goals}




\section{System Design}

\begin{figure}[t]
    \centering
    \includegraphics[scale=0.50]{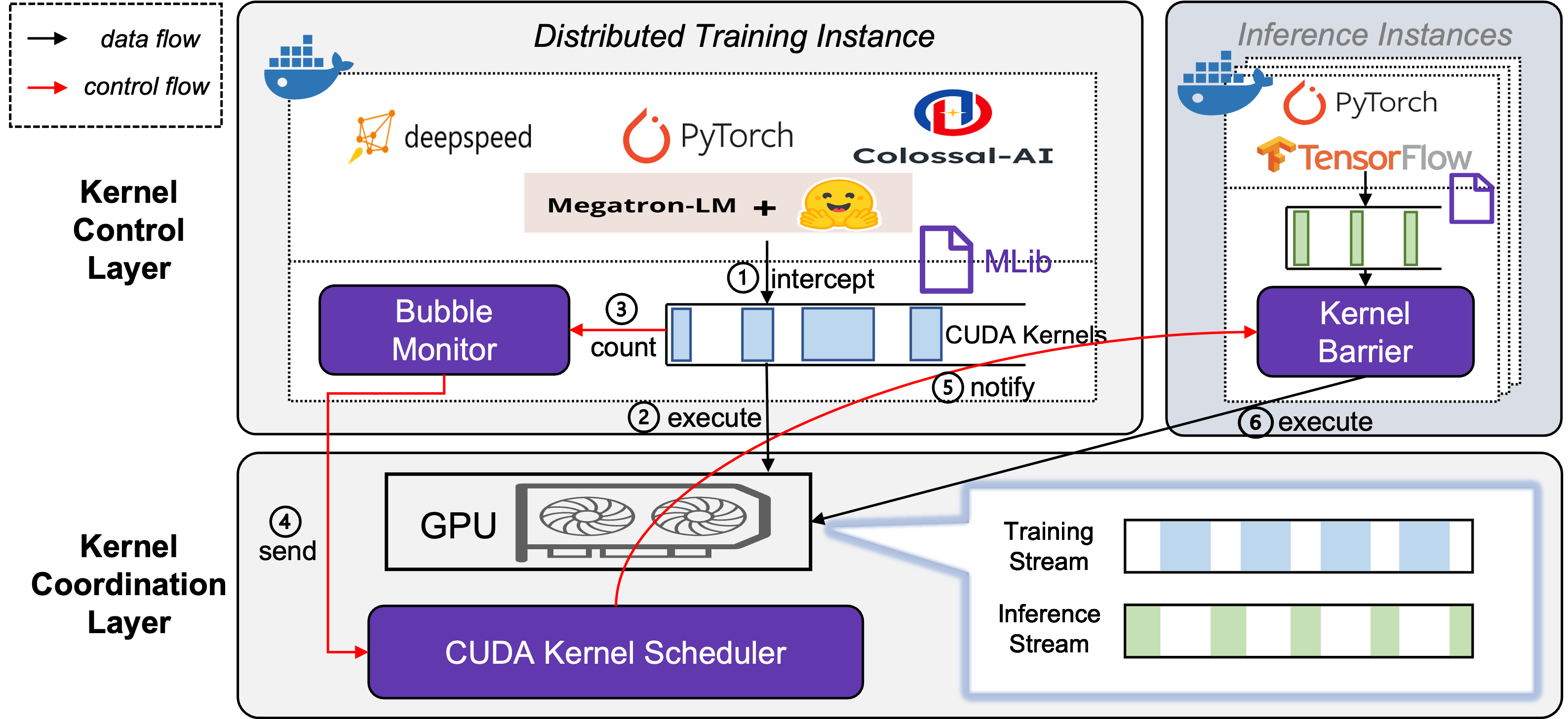}
    \caption{The system architecture of SpecInF. }
    \label{fig:architecture}
\vspace{-0.2in}
\end{figure}

\subsection{Overview}
SpecInF builds on the key idea of speculatively filling inference workloads into bubble periods\footnote{For simplicity, we refer to the GPU-idling period in training mentioned above as bubbles.} produced by distributed training tasks, as illustrated in Figure~\ref{fig:architecture}. It consists of a kernel control layer and a coordination layer.

At the kernel control layer, training and inference workloads can be observed and manipulated in the view of CUDA kernels. First, SpecInF collocates one or more additional inference instances with a training instance to share a GPU, preparing the runtime context to speculatively fill inference workloads.
Second, to observe the GPU utilization of training instances, SpecInF initiates the Bubble Monitor (BM) for each training worker to detect idle periods.
It is suitable for general communication-optimized distributed DL frameworks, such as DeepSpeed \cite{DeepSpeed}, Megatron-LM \cite{shoeybi2019megatron}.
BM intercepts\ding{182} and counts\ding{184} the CUDA kernels, issued at runtime by training instances, which are dispatched to GPU for execution\ding{183}.
Third, SpecInF leverages the Kernel Barrier (KB) to take charge of either blocking or forwarding CUDA kernels of inference instances to the GPU.


At the coordination layer, the CUDA Kernel Scheduler (CKS) on each node is responsible for recognizing training bubbles\ding{185} and deciding speculative inference execution timings of collocated inference instances\ding{186}. SpecInF introduces two distinct mechanisms tailored for offline and online inference tasks, acknowledging their varying degrees of SLO-sensitivity (Service Level Objective), which is explained in Section~\ref{lab:kernel-barrier}.

\subsection{Collocation Basics}

\label{sec:collocation}


Before running the speculative filling workflow, collocation should be done in advance. There are several factors explicitly affecting the collocation policies, including distinct spatial and temporal GPU requirements of tasks (as shown in Section~\ref{sec:opportunity}), risks of OOM errors, training interruptions and throughput degradation due to severe resource contention. According to profilings, SpecInF adheres to the following principles:




\begin{itemize}
    \item \textbf{Principle-I}: The sum of GPU memory occupied at the peak of each collocated instance must be less than the upper limit (e.g., 40GB) of a specific GPU. Based on it, SpecInF collocates inference instances as much as possible to enhance aggregated throughput and GPU utilization.

    \item \textbf{Principle-II}: The minimal execution time (i.e., batch size=1) of collocated online inference must be shorter than the maximal bubble of the primary training task, to at least serve one inference request.
\end{itemize}

\subsection{Speculative Inference Filling}

\subsubsection{Bubble Monitor.}
\label{sec:monitoring}
The bubble monitor detects the possible GPU idling time by observing the CUDA kernel issuing ratio of training instances. A straightforward method to monitor GPU utilization is leveraging tools such as nvml \cite{nvml} at the application level.
However, this approach has two issues. First, the statistical data collection presents a non-negligible delay. The sampling period may vary between 1 and 1/6 seconds, depending on the devices used. Second, the data is updated at fixed intervals, e.g., every 200ms. Thus, the tools may fail to quickly and accurately recognize training bubbles, hindering effective speculative filling.

SpecInF employs a hijacking statistics method. Specifically, it mounts the monitoring library (MLib) to training instances. The MLib intercepts the CUDA kernels and records the counts periodically (e.g., 2ms), which are saved to a sliding window maintained by the BM. Concurrently, the BM calculates the number of continuous zero-counts of sliding windows and then sends it to the CUDA Kernel Scheduler on the node.


\begin{algorithm}[t]
\caption{Adaptive Kernel Scheduling Algorithm}
\footnotesize
\begin{algorithmic}[1]

\State \textbf{Input:}
\State \hspace{\algorithmicindent} $Z_c$: The kernel zero-count from the Bubble Monitor.
\State \hspace{\algorithmicindent} $\alpha, \beta$: The thresholds of two phases.
\State \hspace{\algorithmicindent} $\gamma$: The multiplicative coefficient for offline inference tasks.
\State \hspace{\algorithmicindent} $m $: The number of collocated inference instances.
\State \hspace{\algorithmicindent} $UL, LL $: The upper and lower limit of tokens.

\State \textbf{Output:}
\State \hspace{\algorithmicindent}  $tokens$ for offline inference and $status$ for online, used by speculation execution.

        \If{$Z_c \leq \alpha$}
            \State $tokens \gets 0$, $status \gets busy$ 
            \Comment{conservative phase}
        \ElsIf{$Z_c \leq \beta$}
            \State $tokens \gets \min(LL, tokens * \gamma)/m$,  $status \gets busy $
            \Comment{incremental phase}
        \Else
            \State $tokens \gets \min(UL, tokens * \gamma)/m$,  $status \gets idle $ 
            \Comment{stable phase}
        \EndIf

\end{algorithmic}
\label{algorithm:task_scheduling_resource_limitation}
\end{algorithm}

\subsubsection{CUDA Kernel Scheduler.}
\label{sec:speculation}
The CKS can oversee all GPUs within the node. It acts as a bridge to coordinate the bubble monitoring and speculative filling. It evaluates the GPU busy/idle status information sent by a specific BM and determines the timing and amount of speculative filling. 

The Algorithm~\ref{algorithm:task_scheduling_resource_limitation} outlines the adaptive scheduling logic for a specific GPU. It is divided into three phases: the \textit{conservative phase}, the \textit{incremental phase}, and the \textit{stable phase}. During the conservative phase (line 9-10), CKS maintains the GPU \textit{logically} busy to prevent interfering with ongoing CUDA execution of relatively longer training kernels. In the incremental phase (line 11-12), CKS opts to gradually increase tokens allocated to each offline inference instance, instead of maximizing them immediately, to mitigate interference associated with asynchronous execution. The \textit{busy} status setting for online inference stems from the same reason. In the stable phase (lines 13-14), CKS capitalizes on a relatively steady bubble period, typically characterized by intense communication workloads, to maximize GPU utilization.


\subsubsection{Kernel Barrier.} 
\label{lab:kernel-barrier}
Both online and offline inference workloads can be used to fill training bubbles by the KB, while online workloads need explicit SLO guarantees. 
For \textit{offline inference} workloads, the KB receives allocated tokens from the CKS to serve corresponding inference workloads. Each kernel, when forwarded to the GPU, consumes tokens proportionate to its size. It also collects CUDA kernel count information, similar to the BM. If the remaining tokens during a given period are insufficient, the KB blocks subsequent inference kernel issuing in the queue. Conversely, when collocated training resumes, the tokens sent by CKS may decrease to zero. 
It is crucial to note that despite the asynchronous nature of CUDA kernel execution, our observations via the Nsight System indicate that for inference tasks, the CUDA APIs issued by CPU are almost synchronously triggered. The preceding CUDA API waits for its corresponding kernel to complete on the GPU before launching the next one, with most kernel execution times under 1 ms.  In summary, the token release and block mechanism effectively minimizes resource contention with collocated training, thereby ensuring throughput.


For \textit{online inference} workloads, KB additionally adopts a real-time pull-and-execute mechanism to meet the demands of stringent SLOs, since it is challenging to fill online requests into training bubbles without mutual interference due to the unpredictability of workloads. In this case, KB proactively pulls online requests one by one from the request queue, upon receiving the \textit{idle} signal (set by Algorithm~\ref{algorithm:task_scheduling_resource_limitation}). Based on Principle-II in Section~\ref{sec:collocation}, SpecInF ensures to handle at least one request during bubbles of each training iteration. Moreover, to avoid training resuming immediately after pulling one request, CKS preemptively sets the status to \textit{busy}, according to profiling information on training iteration time (e.g., 1.5 seconds). During the \textit{busy} status, requests are handled by other inference instances.  

\section{Implementation}
We implement a prototype of SpecInF with 2k C LOCs and 2k Python LOCs for evaluations. In the prototype, all training and inference tasks run with \textit{NVIDIA-docker} runtime.

The MLib is based on Linux LD\_PRELOAD mechanism, where the interception libraries are written in /etc/ld.so.preload file, allowing the hook logic to be loaded before the standard CUDA APIs. The BM is running as a \textit{pthread} within the MLib process. The KB follows the same implementation logic as MLib, while the difference is that it sets up a \textit{pthread} to forward or block the inference kernels. The CUDA Kernel Scheduler, running as a daemon, actively establishes a \textit{UNIX socket} with each collocated instance to receive or send information.



\section{Evaluation}

\subsection{Methodology}
\textbf{Experiment testbed}. We evaluate SpecInF on a GPU server with 4 * NVIDIA A100-40GB, equipped with PyTorch v1.11, DeepSpeed v0.11.1, CUDA v11.7.


\textbf{Workloads}. For the training workloads, we adopt BERT-base and RoBERTa-large training based on Pytorch.DDP for DP mode, LLaMA2-7B and ChatGLM-6B fine-tuning with DeepSpeed for MP and PP modes.
For inference workloads, we employ medium-sized models, such as Resnet152, BERT-base, VGG19, RoBERTa-large, and GPT2-large.
Poisson distribution \cite{strati2024orion} is used for generating online inference workloads. For the collocation cases of RoBERTa-Resnet and RoBERTa-VGG in section~\ref{exp:speculative-exec}, the mean value is set to 30. In other scenarios, we use a mean value of 10 across 2000 total requests.



\textbf{Metrics}. For evaluating distributed training and offline inference, the primary metrics include tokens per second (tokens/s) for NLP models and samples per second (samples/s) for CV models. To facilitate direct comparisons, we normalize these metrics to throughputs without GPU sharing. For online inference, we focus on tail latency (i.e., p95). 

\textbf{Baselines}. We compare SpecInF with the following methods.
\begin{itemize}
    \item \textbf{MPS} \cite{NVIDIAMPS}. A popular spatial GPU sharing technique developed by NVIDIA. It is used by many works \cite{yang2022infless,gu2023fast}.
    \item \textbf{TGS} \cite{wu2023transparent}. A transparent GPU sharing mechanism between DL jobs, focuses on guaranteeing productive job throughputs. 
    \item \textbf{Co-Exec}. The strawman GPU sharing method despite resource contention.
    \item \textbf{Exclusive}. Each training or inference instance monopolizes the whole GPU.
\end{itemize}

\begin{figure*}[t] 
    \centering
    \setlength{\abovecaptionskip}{-0.cm}
    \subfigure[Offline performance]{
            \includegraphics[scale=0.77]{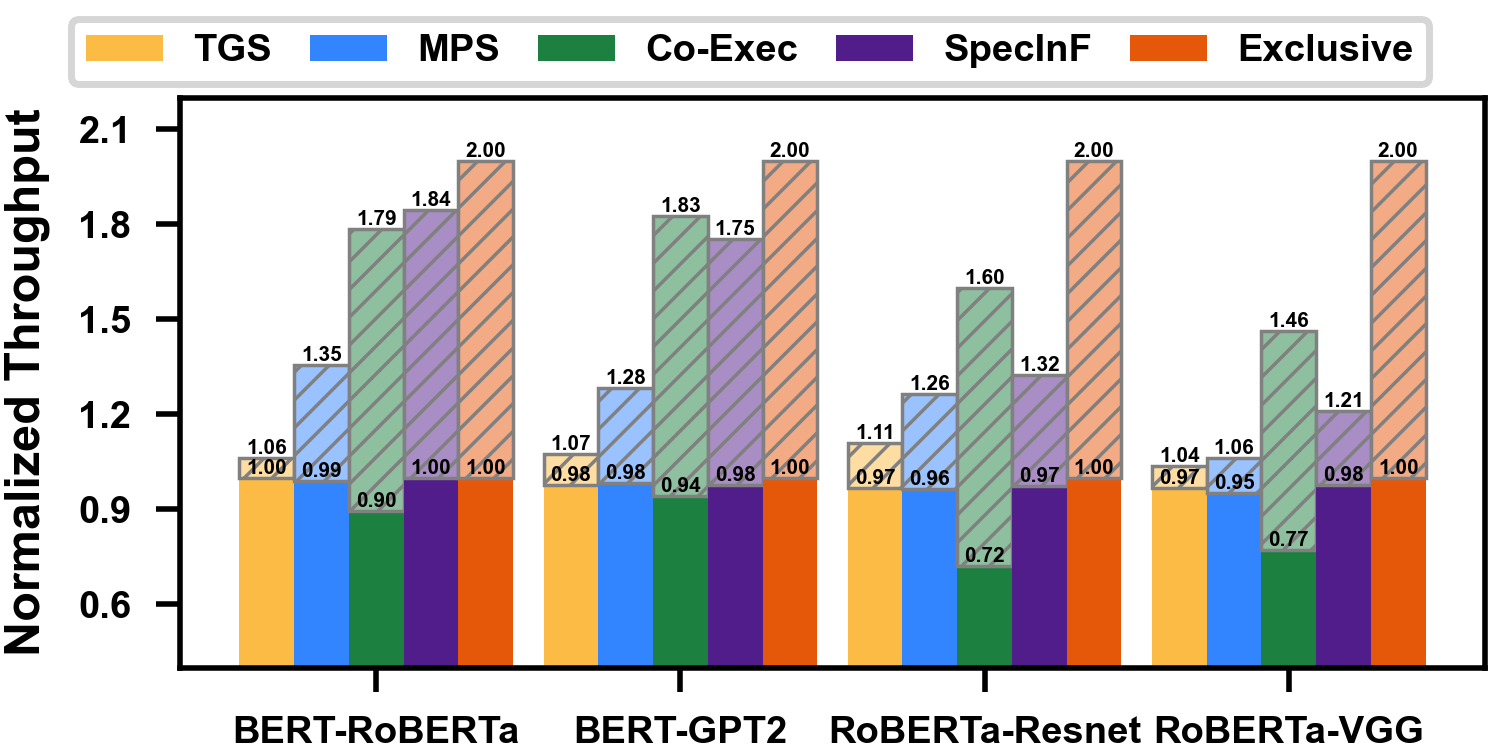}
		\label{fig:ddp-offline}
    }
    \hspace{-0.1in}
    \subfigure[Online performance]{
     	\includegraphics[scale=0.77]{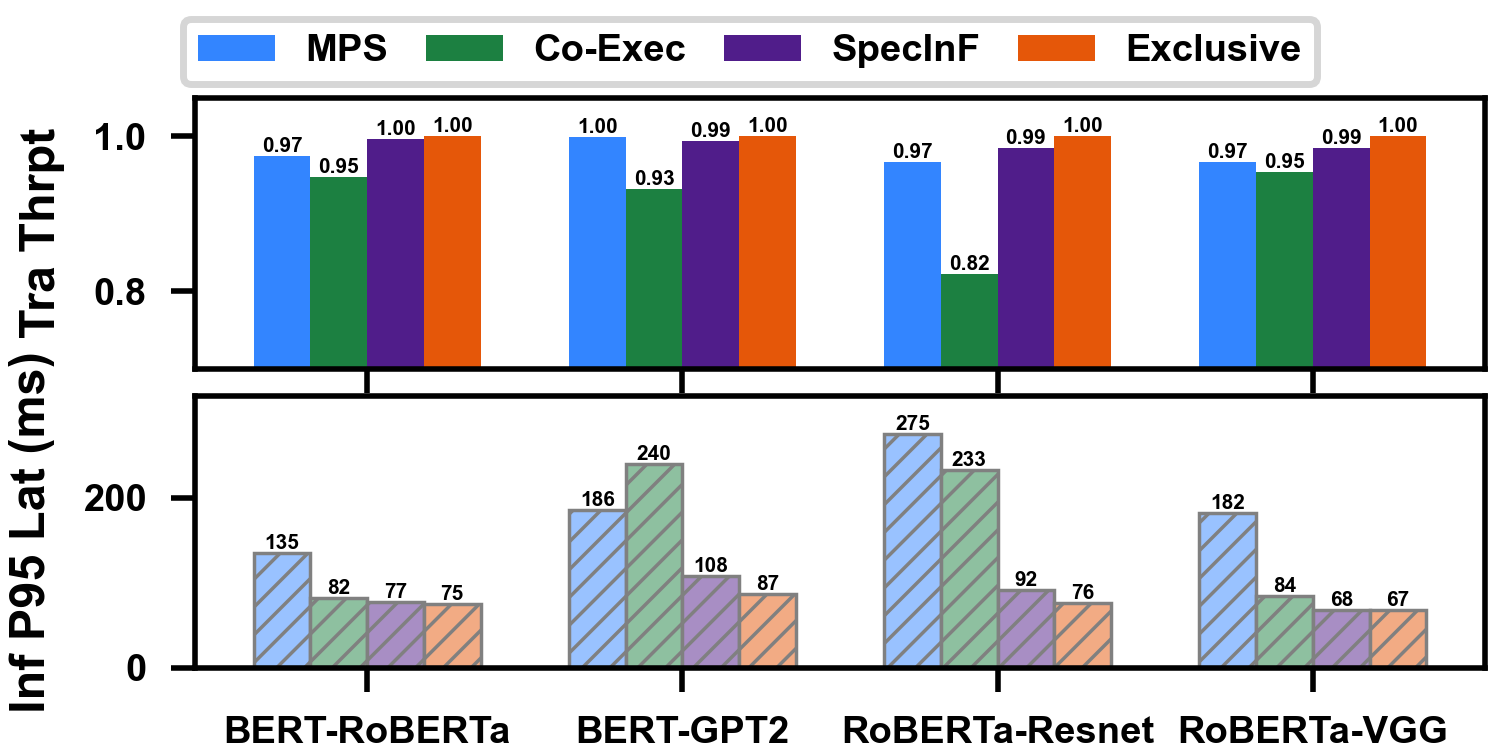}
            \label{fig:ddp-online}
    }

\caption{DP performance comparison: (a) the solid bar represents normalized training throughput and the light bar with dashed lines represents normalized offline inference throughput. (b) bars in the upper subfigure indicate normalized training throughputs, and the lower subfigure shows p95 latency of online inference. TGS is excluded due to excessive tail latencies.} 
	\label{fig:ddp-performance}
\vspace{-0.1in}
\end{figure*}

\begin{figure*}[t] 
    \centering
    \setlength{\abovecaptionskip}{-0.cm}
    \subfigure[Offline performance]{
            \includegraphics[scale=0.77]{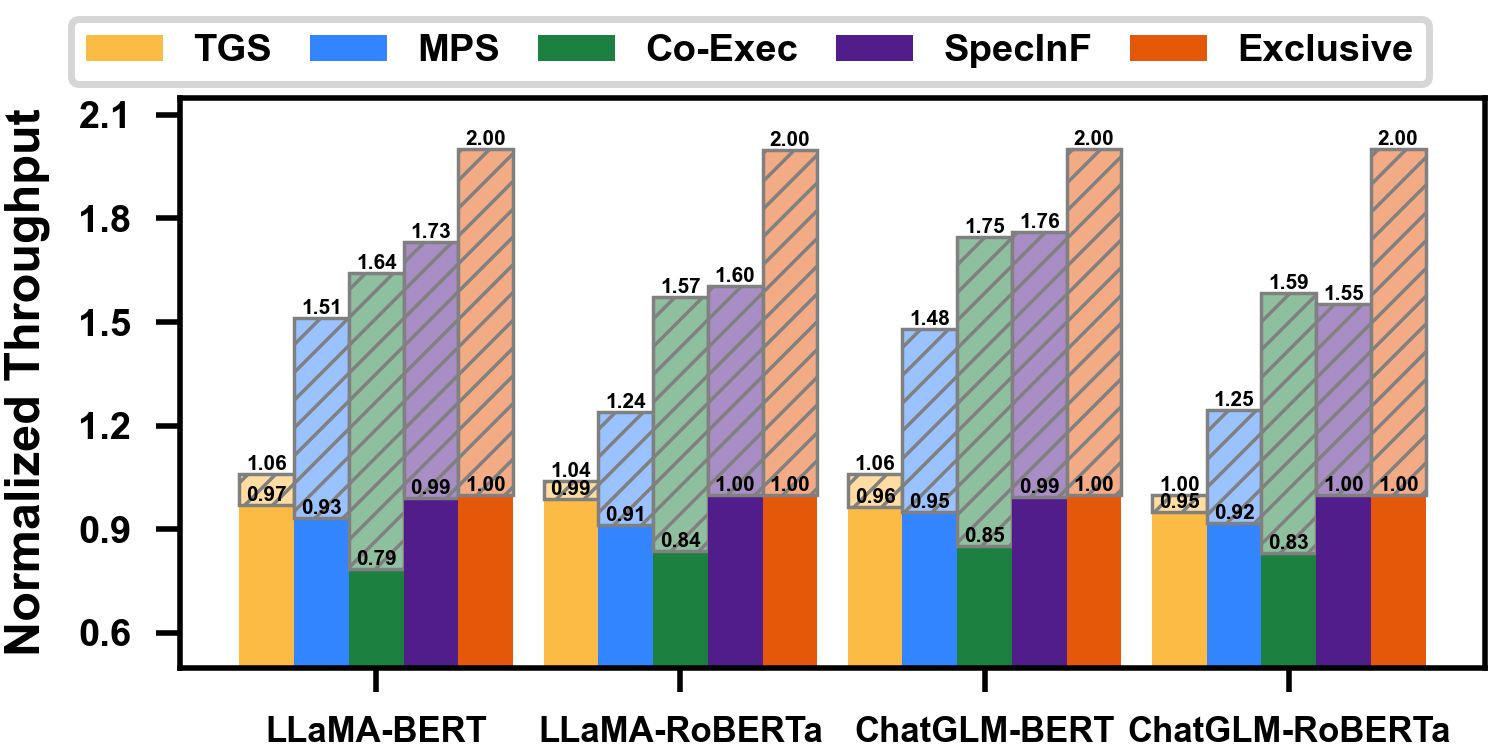}
		\label{fig:mp-offline}
    }
    \hspace{-0.1in}
    \subfigure[Online performance]{
     	\includegraphics[scale=0.77]{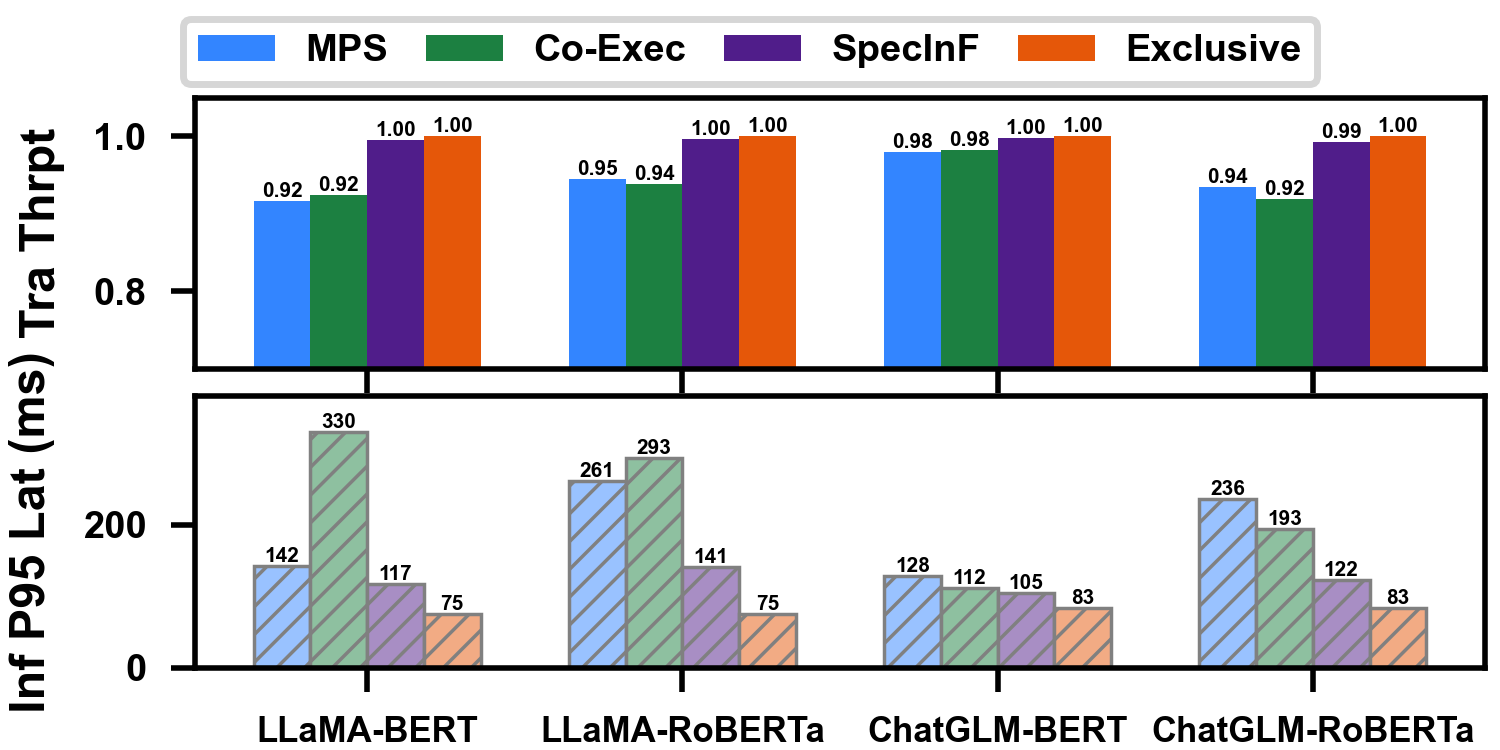}
            \label{fig:mp-online}
    }
	
\caption{MP performance comparison.}
	\label{fig:mp-performance}
\vspace{-0.2in}
\end{figure*}

\begin{figure*}[t] 
    \centering
    \setlength{\abovecaptionskip}{-0.cm}
    \subfigure[Offline performance]{
            \includegraphics[scale=0.77]{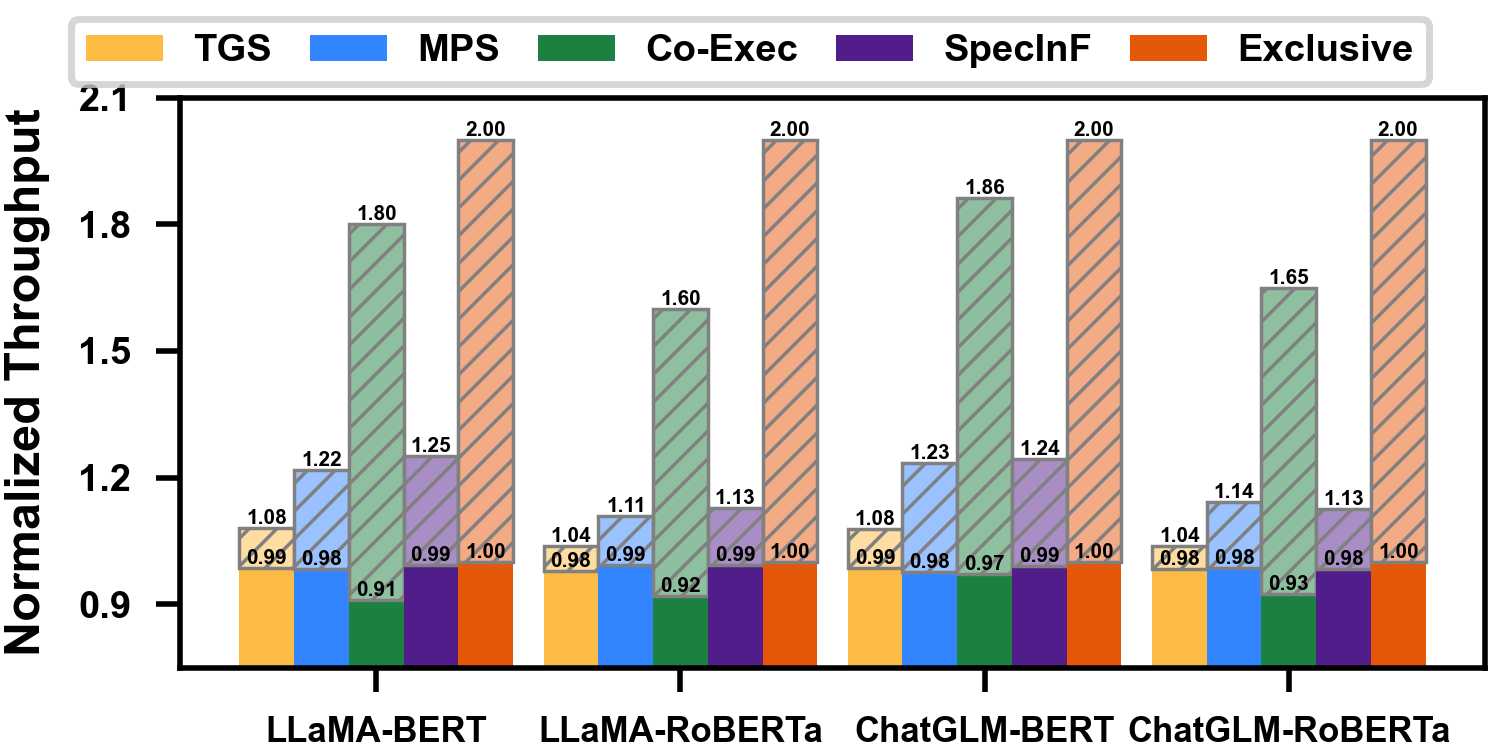}
		\label{fig:pp-offline}
    }
    \hspace{-0.1in}
    \subfigure[Online performance]{
     	\includegraphics[scale=0.77]{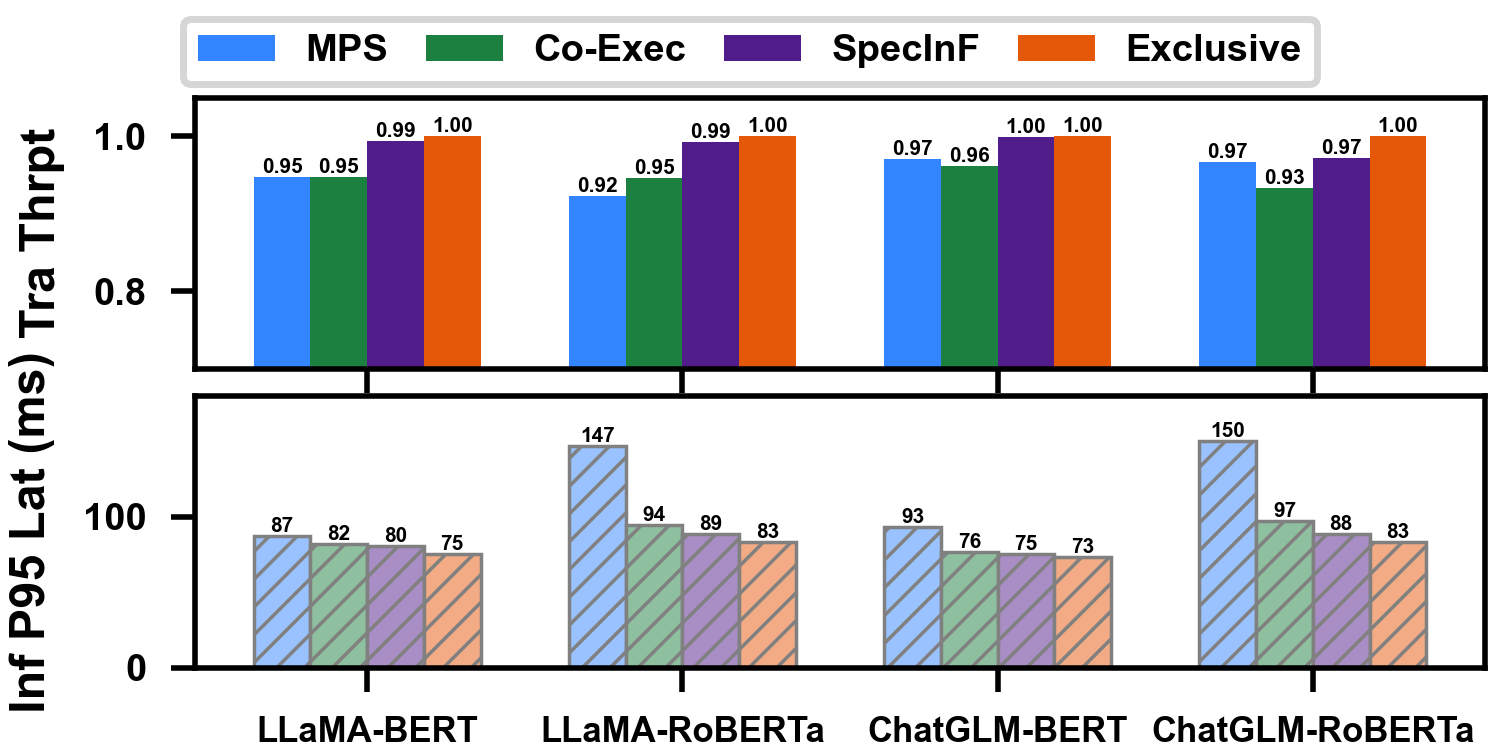}
            \label{fig:pp-online}
    }
	
\caption{PP performance comparison.}
	\label{fig:pp-performance}
\vspace{-0.2in}
\end{figure*}

\subsection{Speculative Inference Filling Performance}
\label{exp:speculative-exec}

\textbf{Offline Inference Filling}. Figures~\ref{fig:ddp-offline},\ref{fig:mp-offline} and \ref{fig:pp-offline}  demonstrate that SpecInF delivers high throughputs for offline inferences with training throughput guarantees in DP, MP and PP modes.
Considering the primary training workload, all baselines, except Co-Exec, generally maintain the performance. 
For the collocated offline inference workloads, SpecInF provides 23-84\% throughput of Exclusive and 33-94\% of Co-Exec, best in other baselines. However, the Exclusive requires one additional GPU, and Co-Exec significantly reduces collocated training throughput (e.g., up to 28\% in the RoBERTa-Resnet case), failing to meet the goals in Section~\ref{Sec:goals}.

In DP cases, inference throughput is 1.23$\times$-3.5$\times$ of MPS, and 2.9$\times$-14$\times$ of TGS. The underperformance of TGS is mainly due to its inadequate bubble detection and relatively conservative time-share strategy, while MPS statically limits GPU resources available to the inference process.
In MP cases, SpecInF achieves the highest aggregated throughputs in the first three cases with training throughput guarantees. Specifically, the inference throughput increases by up to 80\% and 11$\times$ compared to MPS and TGS, respectively. 
In PP cases, the advantages of SpecInF become relatively marginal to DP and MP cases. Co-Exec achieves significantly higher inference throughput but sacrifices 3\%-9\% of training throughput. SpecInF's performance is comparable to MPS and superior to TGS. The underlying reason is that though dividing a mini-batch into small micro-batches shortens bubbles, it leaves GPU consistently underutilized, thus allowing it to sufficiently handle inference workloads in Co-Exec mode. In the future, for the PP scenario, we aim to execute inference workloads concurrently as much as possible until the BM observes that training iterations have lengthened.

\textbf{Online Inference Filling}. Figures ~\ref{fig:ddp-online},\ref{fig:mp-online},\ref{fig:pp-online} demonstrate that SpecInF can deliver low p95 latencies of online inference with training throughput guarantees.
SpecInF consistently shows the lowest p95 latency of inferences, trailing only behind the Exclusive mode, while maintaining standard training throughputs.
This advantage is owing to the proactive pull-and-execute mechanism.
In DP cases, shown in Figure~\ref{fig:ddp-online}, SpecInF reduces p95 by up to 61\% and 67\% compared to Co-Exec and MPS, respectively.
In MP cases, it lowers p95 by an average of 40\% compared to Co-Exec and 33\% compared to MPS.
Similar to offline cases above, the gains in PP modes diminish, but SpecInF still maintains the best tail latency performance except Exclusive. 

\subsection{Multi-instance Support}
\begin{figure*}[t] 
    \centering
    \setlength{\abovecaptionskip}{-0.cm}
    \subfigure[DP performance]{
            \includegraphics[scale=0.77]{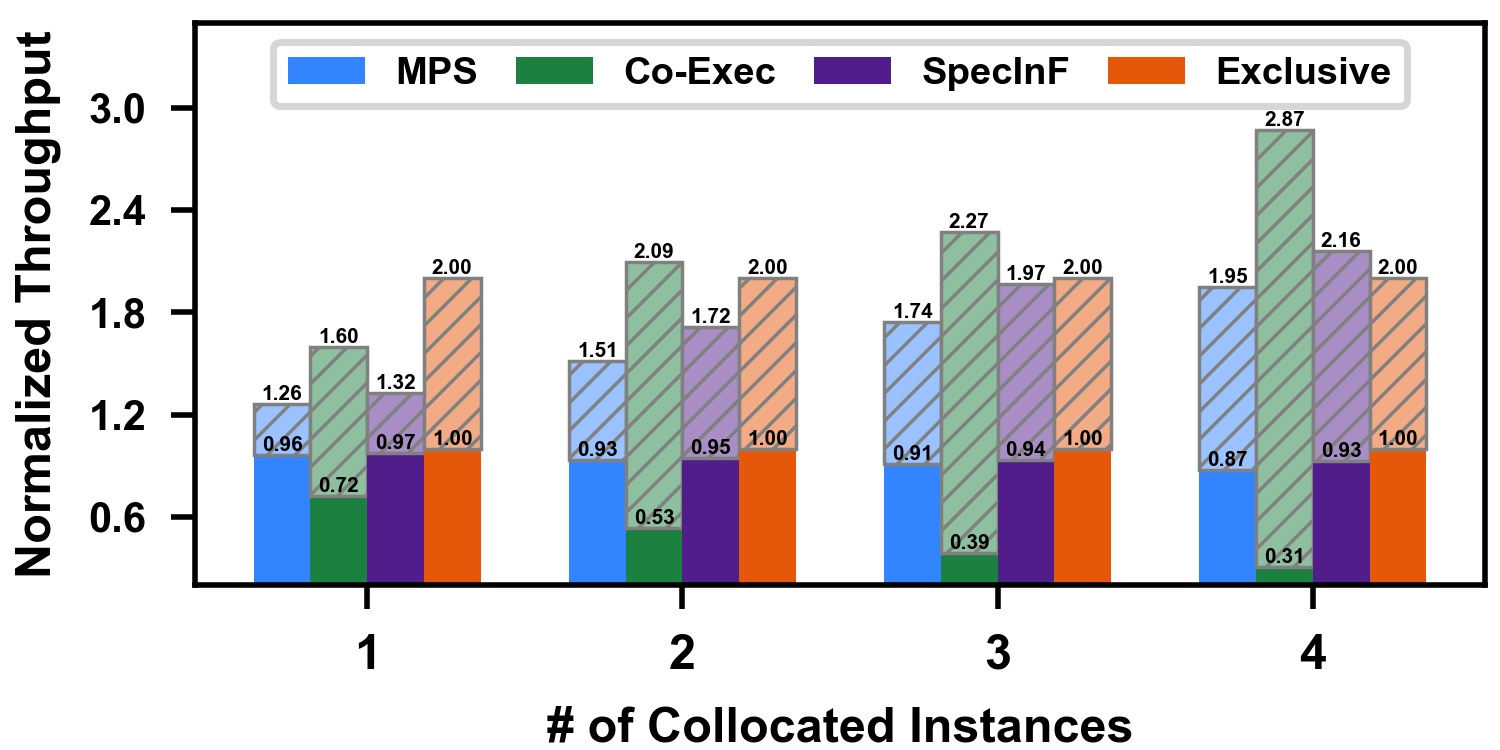}
		\label{fig:ddp-varysize}
    }
    \hspace{-0.1in}
    \subfigure[MP performance]{
     	\includegraphics[scale=0.77]{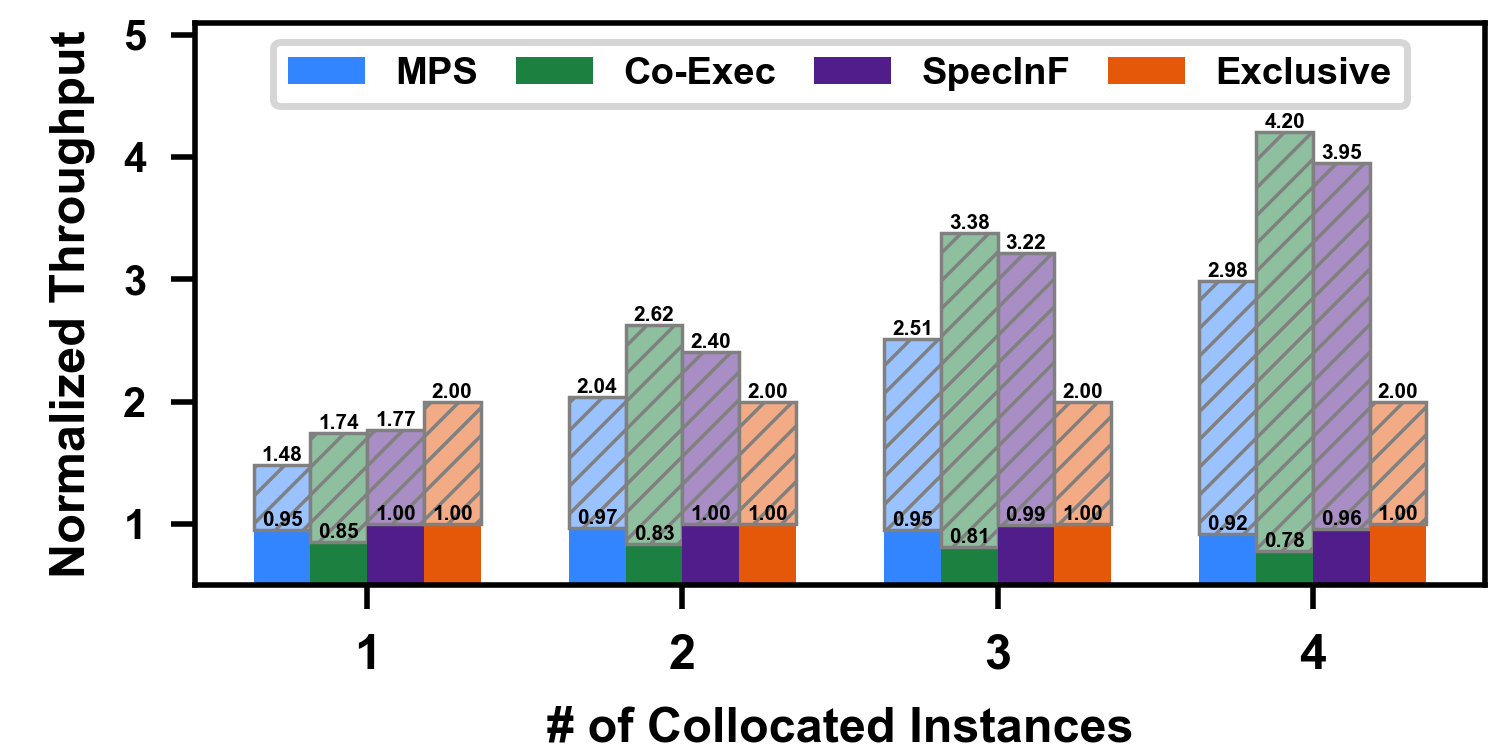}
            \label{fig:mp-varysize}
    }
\caption{Performance comparison under different sizes of collocated inference instances. Here we regard the single instance throughput on the exclusive GPU as the normalized one. (a) the RoBERTa-Resnet collocation case; (b) the ChatGLM-BERT collocation case. }
	\label{fig:varysize-performance}
\vspace{-0.2in}
\end{figure*}

\textbf{Aggregated Throughput Improvement with Multiple Inference Instances Support}. As Principle-I in Section~\ref{sec:collocation} mentions, SpecInF supports collocating multiple inference instances to enhance GPU utilization. Figure~\ref{fig:varysize-performance} shows that SpecInF achieves a sub-linear growth in inference throughput while ensuring training throughput, with increasing collocated inference instances.
In DP cases depicted in Figure~\ref{fig:ddp-varysize}, SpecInF outperforms Exclusive by achieving an additional 35\%-123\% in inference throughput when the number of instances ranges from 1 to 4, with a maximum training throughput reduction of less than 7\%.
Although Co-Exec surpasses all other baselines in aggregated throughput, it leads to a substantial degradation in training, up to 61\%.
Consistent trends are observed in the MP scenario shown in Figure~\ref{fig:mp-varysize}.
Specifically, SpecInF matches the offline throughput performance of Co-Exec while avoiding the latter's detrimental impact on training, which can reach up to 22\%. Notably, with 4 instances, the inference throughput soars to 299\% more than that of Exclusive.

\subsection{System Overhead}



\begin{wrapfigure}{l}{7cm}
    \setlength{\abovecaptionskip}{0pt} 
    \vspace{-2\baselineskip}
    \includegraphics[scale=0.9]{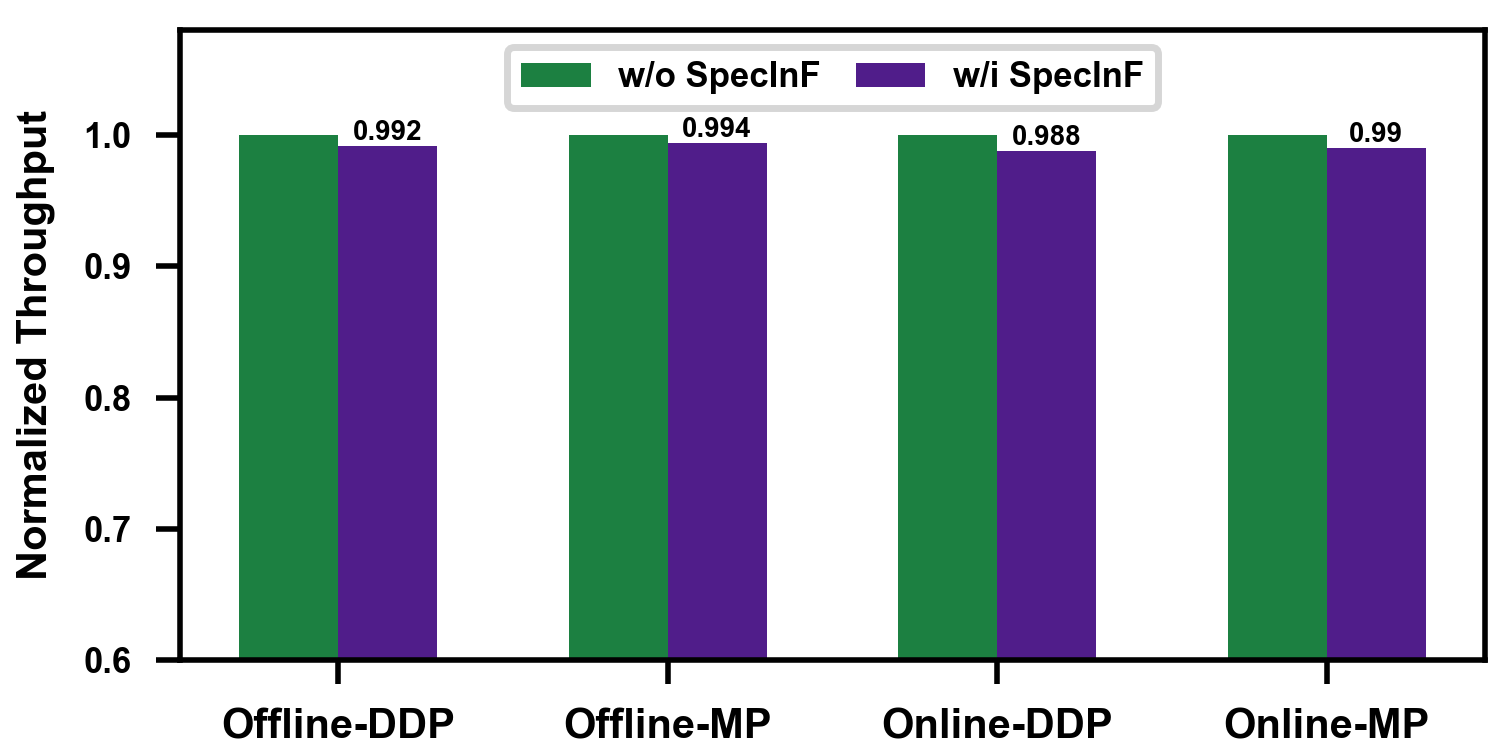}
    \caption{The system overhead of SpecInF.}
    \label{fig:overhead}
     \vspace{-0.2in}
\end{wrapfigure}

\textbf{Negligible System Overhead.} We evaluate the system overhead by collocating the training instance with inference instances, but without triggering any inference requests. These scenarios include BERT-RoBERTa in Figure~\ref{fig:ddp-performance} and ChatGLM-RoBERTa in Figure~\ref{fig:mp-performance}). The results shown in Figure~\ref{fig:overhead} indicate the overheads of speculative filling including bubble monitoring, kernel scheduling and kernel barrier, are minimal (i.e., 1\%), which is considered acceptable.



\section{Related Work}

\textbf{DL Scheduling}. As DL evolves, systems adopt various scheduling methods. Tiresias \cite{gu2019tiresias} provides elastic training, without explicit resource scheduling, to improve throughput or reduce Job Completion Time.
Works like \cite{yang2022infless,gu2023fast} focus on supporting GPU-efficient and high-throughput inference serving. Mixing them up, Lyra \cite{li2023lyra} loan inference GPUs to train models in the long run (i.e., at hours level). Orion \cite{strati2024orion} collocates small-sized training and inference tasks but at the thread level, which is not suitable for cloud containers.


\textbf{GPU Sharing}. The straightforward method to enhance GPU utilization effectively is to share a single GPU with multiple DL tasks. Existing works can be categorized into temporal and spatial sharing approaches. For temporal sharing, Antman \cite{xiao2020antman} implicitly inserts training jobs during iterations. Based on this, TGS \cite{wu2023transparent} provides high transparency.  However, both methods are not suitable for LLMs due to large memory footprints. As for spatial sharing, MIG \cite{NVIDIAMIG} supports the physical isolation of GPU devices but lacks adaptability. MPS \cite{NVIDIAMPS} is widely used in DL systems \cite{gu2023fast,yang2022infless}, while it can not dynamically consume idle compute resources completely due to static allocation. Aiming at distributed training, SpecInF detects bubbles timely and speculatively fills inference workloads to improve GPU utilization.

\section{Conclusion}

Nowadays, Deep Learning revolutionizes various aspects of life.
However, GPUs used for training these DL applications are usually underutilized, yielding massive compute and memory fragments.
We observe that moderate inference workloads are well-suited to fill up these GPU fragmentations.
In this paper, we present SpecInF, which collocates distributed training with online/offline inference instances, to speculatively serve inference workloads, significantly improving GPU utilization. The results show that SpecInF can exploit the idling GPU resources in various distributed training modes, delivering additional up to 14$\times$ offline inference throughputs than TGS and 67\% reduction in online inference p95 latency than MPS, while guaranteeing collocated training throughput.

\section*{Acknowledgement}
We thank our colleague Xiaohong Wang for her kind support in this study. This work is supported by the Innovation Funding of ICT, CAS under Grant No. E461040, No. E361060, and the Pilot for Major Scientific Research Facility of Jiangsu Province of China under Grant No. BM2021800.








\normalem
\bibliographystyle{plain}  
\bibliography{main}  

\end{document}